\documentclass[copyright,creativecommons]{eptcs}
\providecommand{\event}{PLACES 2015} 

\usepackage{breakurl}
\usepackage{stmaryrd}
\usepackage{latexsym}
\usepackage{listings} 
\usepackage{color, colortbl}
\usepackage{stmaryrd}
\usepackage{latexsym}
\usepackage{amssymb}
\usepackage[leqno,tbtags]{amsmath}
\usepackage{xcolor}
\usepackage{multicol}
\usepackage{amsthm}

\newcommand{\alt}{~|~}
\newcommand{\parc}[2]{#1 \,\|\, #2}
\newcommand{\send}[2]{\textsf{send}~#1~#2}
\newcommand{\sendu}[2]{\underline{\textsf{send}}~#1~#2}
\newcommand{\stable}[1]{\textsf{stable}~#1}
\newcommand{\stableu}[1]{\underline{\textsf{stable}}~#1}
\newcommand{\backtrack}{\textsf{backtrack}}

\newcommand{\recvtt}{\textsf{recv}}

\newcommand{\recvo}[3]{\recvtt ~#2(#1).#3}

\newcommand{\hllproc}[4]{\langle #1@#2:~#3, #4 \rangle}

\newcommand{\diff}[1]{\textcolor{red}{#1}}
\newcommand{\salmodel}{refine_2pb.sal}
\newcommand{\HP}{HP}
\newcommand{\LP}{LP}

\theoremstyle{remark}
\newtheorem{trans}{Trans}

\lstdefinelanguage{scala}{morekeywords={class,object,trait,extend,with,new,if,while,for,def,val,var,this},
otherkeywords={->,=>},
sensitive=true,
morecomment=[l]{//},
morecomment=[s]{/*}{*/},
morestring=[b]''}

\title{Reversible Communicating Processes}

\author{
Geoffrey Brown
\qquad\qquad
Amr Sabry
\institute{School of Informatics and Computing\\ 
Indiana University, Bloomington, IN}
\email{\quad geobrown@indiana.edu \quad\qquad sabry@indiana.edu}
}

\def\titlerunning{Reversible Communicating Processes}  
\def\authorrunning{Brown \& Sabry}

\begin{document}
\maketitle

\begin{abstract}

  Reversible distributed programs have the ability to abort
  unproductive computation paths and backtrack, while unwinding
  communication that occurred in the aborted paths.  While it is
  natural to assume that reversibility implies full state recovery (as
  with traditional roll-back recovery protocols), an interesting
  alternative is to separate backtracking from local state recovery.
  For example, such a model could be used to create complex
  transactions out of nested compensable transactions where a
  programmer-supplied compensation defines the work required to
  ``unwind'' a transaction.

  Reversible distributed computing has received considerable
  theoretical attention, but little reduction to practice; the few
  published implementations of languages supporting reversibility
  depend upon a high degree of central control.  The objective of this
  paper is to demonstrate that a practical reversible distributed
  language can be efficiently implemented in a fully distributed
  manner.

  We discuss such a language, supporting CSP-style synchronous
  communication, embedded in Scala. While this language provided the
  motivation for the work described in this paper, our focus is upon
  the distributed implementation.  In particular, we demonstrate that
  a ``high-level'' semantic model can be implemented using a simple
  point-to-point protocol. 
  \vspace{-1em}

\end{abstract}

\section{Introduction}

Speculative execution either by intent or through misfortune (in
response to error conditions) is pervasive in system design and yet it
remains difficult to handle at the program
level~\cite{springerlink:10.1007/978-3-642-15763-9}.
Indeed, we find that despite the importance of speculative
computation, there is very little programmatic support for it in
distributed languages at the foundational level it deserves. We note
that, from a programming language perspective, speculative execution
requires a backtracking mechanism and that, even in the sequential
case, backtracking in the presence of various computational effects
(e.g.\ assignments, exceptions, etc.) has significant
subtleties~\cite{Hinze:2000:DBM:351240.351258}. The introduction of
concurrency additionally requires a ``distributed backtracking''
algorithm that must ``undo'' the effects of any communication events
that occurred in the scope over which we wish to backtrack. While this
has been successfully accomplished at the algorithmic level (e.g.\ in
virtual time based
simulation~\cite{jefferson85,Jefferson:1987:TWO:41457.37508}), in
models of concurrent languages (e.g.\
\cite{Danos:2005:TR:1099332.1099364,Danos:2007:ST:1243510.1243684,
  Lanese:2011:CRH:2040235.2040261,reversingpi,Phillips:2007:RMC:1298755.1298827})
and in some restricted parallel shared-memory environments
(e.g.~\cite{deVries:2010,springerlink:10.1007/BF03037164,Lesani:2011,cml,muoz}),
it does not appear that any concurrent languages based upon message
passing have directly supported backtracking with no restrictions.
The language constructs we introduce are inspired by the stabilizers
of \cite{stabilizers}; however, that work depends upon central control
to manage backtracking.  Our work was also inspired by the work of
Hoare and others \cite{LiZhuHe2008,Hoare2010}. Communicating message
transactions \cite{deVries:2010,Lesani:2011} is an interesting related
approach that relies upon global shared data structures.



The work presented in this paper has a natural relationship to the
rich history of rollback-recovery
protocols~\cite{Elnozahy:2002:SRP:568522.568525}.  Rollback-recovery
protocols were developed to handle the (presumably rare) situation
where a processor fails and it is necessary to restart a computation
from a previously saved state.  The fundamental requirement of these
protocols is that the behavior is as if no error ever occurred.  In
contrast, we are interested in systems where backtracking might take
the computation in a new direction based upon state information
gleaned from an abandoned execution path; the (possibly frequent)
decision to backtrack is entirely under program control.  Because
check-pointing in traditional rollback-recovery protocols involves
saving a complete snapshot of a process's state, it is a relatively
expensive operation. Much of the research in rollback-recovery
protocols focuses upon minimizing these costs.
The cost of check-pointing is much lower for our domain -- saving
control state is no more expensive than for a conventional exception
handler; the amount of data state preserved is program dependent.



Implementing a reversible concurrent language is not a trivial
undertaking and, as we found, there are many opportunities to
introduce subtle errors. Ideally, such a language implementation
should be accompanied by a suitable semantics that provides both a
high-level view which a programmer can use to understand the expected
behavior of a program text, and a low-level view which the language
implementer can use to develop a correct implementation. In order to
accommodate these two constituencies, we have developed two separate
semantic models.  We have developed a refinement mapping to
demonstrate that the low-level model is a correct implementation of
the high-level model.


The remainder of this paper is organized as follows.  We begin with a
small example that illustrates the main ideas using our Scala
implementation. We then, in Sec.~\ref{sec:hlsemantics}, present a
formal ``high-level'' semantic model focusing on the semantics of
forward communication and backtracking. In
Sec.~\ref{sec:chanprotocol}, we discuss a communication protocol for
maintaining backtracking state across distributed communicating
agents. Sec.~\ref{sec:ll} introduces a ``low-level''
model that utilizes the channel protocol to implement the high-level
model, and outlines the proof of correctness of the low-level
semantics with respect to the high-level semantics. We end with a
brief discussion.

\section{Example}
\label{sec:examples}

A programmer wishing to use our distributed reversible extensions of
Scala imports our libraries for processes and channels and then
defines extensions of the base class \lstinline+CspProc+ by overriding the
method \lstinline+uCode+. The user-defined code must use our channel
implementation for communication and may additionally use the keywords
\lstinline+stable+ and \lstinline+backtrack+ for managing backtracking over
speculative executions. The \lstinline+stable+ regions denote the scope of
saved contexts; executing \lstinline+backtrack+ within a stable region
returns control to the beginning of the stable region -- much like
throwing an exception, but with the additional effect of unwinding any
communication that may have occurred within the stable region.

\begin{figure}[t]
\begin{multicols}{2}
\begin{lstlisting}[numbers=left,firstline=7,lastline=42]
  class p1 (c: SndPort, name: String)
      extends CspProc(name) {
    override def uCode = {
      println("p1: start")
      var count = 2
      stable {
        println("p1: snd " + count)
        send(c,count)
        stable {
          println("p1: snd " + count)
          send(c,count)
          count = count - 1
          if (count > 0) {
            println ("p1: backtrack")
            backtrack }
        }}
    }}

  class p2 (c: RcvPort, name: String)
      extends CspProc(name) {
    override def uCode = {
      println("p2: start")
      stable { 
        var x = receive(c)
        println("p2: recv " + x)
        var y = receive(c)
        println("p2: recv " + y)
      }
    }}

  class RootClass() extends CspProc("root") {
    override def uCode() = {
      stable{
        var c = newChannel("c")
        par (new p1(c.tx, "p1"),
             new p2(c.rx, "p2"))
\end{lstlisting}
\end{multicols}
\caption{\label{fig:ex}Example}
\end{figure}

The excerpt of Fig.~\ref{fig:ex} provides the code for two processes
\lstinline|p1| and \lstinline|p2| that communicate over
channel~\lstinline|c| -- not shown is the code that creates these
processes and the channel. Process \lstinline|p1|'s execution consists
of entering a stable region, entering a nested stable region (line 6),
sending a message to \lstinline|p2| (line 8), entering another stable
region (line 9), sending another message to \lstinline|p2| (line 11),
and then possibly backtracking (line 15). Meanwhile process
\lstinline|p2| also starts a stable region (line 24) in which it
receives the two messages (lines 26 and 29).

A possible execution trace showing a possible interleaving of the
execution is:

\begin{lstlisting}
p2: start
p1: start
p1: snd 2
p2: recv 2
p1: snd 2
p2: recv 2
p1: backtrack
p2: start
p1: snd 1
p2: recv 1
p1: snd 1
p2: recv 1
\end{lstlisting}

\noindent In the trace, processes \lstinline|p1| and \lstinline|p2|
both start executing and enter their respective stable
regions. Process~\lstinline|p2| must block until a communication event
occurs on channel \lstinline|c|. Process \lstinline|p1| initiates the
communication sending the value \lstinline|2| which is received by
\lstinline|p2| which must then block again waiting to receive on
channel~\lstinline|c|. Process \lstinline|p1| enters its nested stable
region and sends another value \lstinline|2| which is received. At
this point, process \lstinline|p2| is ``done'' but
process~\lstinline|p1| decides to backtrack. As a result, process
\lstinline|p1| transfers its control to the inner stable region (line
9). This jump invalidates the communication on channel \lstinline|c|
at line 11. Process~\lstinline|p1| then blocks until process
\lstinline|p2| takes action. When process \lstinline|p2| notices that
the second communication event within the stable region (line 29) was
invalidated, it backtracks to the start of its stable region. This
jump invalidates the first communication action (line 26) which in
turn invalidates the corresponding action in~\lstinline|p1| at line
8. In other words, process \lstinline|p1| is forced to backtrack to
its outer stable region to establish a causally consistent state. It
is important to remember that all processes have an implicit stable
region that includes their full code body; \lstinline|p2| is therefore
forced to backtrack to the beginning of its code.  In other words,
after the backtracking of \lstinline|p1|, both communication events
between \lstinline|p1| and \lstinline|p2| are re-executed.

\section{High-Level Semantics of a Reversible Process Language}
\label{sec:hlsemantics}

We will present two semantic models for our language. The first
``high-level'' semantics formalizes both the forward communication
events that occur under ``normal'' program execution and the backwards
communication events that occur when processes are backtracking to
previously saved states as atomic steps.
In the low-level semantics, these communication events are further
subdivided into actions that communicating senders and receivers may
take independently in a distributed environment and hence trades
additional complexity for a specification that is close to a direct
implementation.  This low-level semantics is based upon a channel
protocol that we have verified using the SAL infinite-state model
checker~\cite{Moura04,Moura03,ds04}; the invariants validated using
SAL were necessary to prove that the low-level semantics is a
refinement of the high level semantics. Both of our semantic models
are based upon virtual time -- a commonly used technique for
conventional rollback recovery
protocols~\cite{fidge1988,mattern88}. Our approach differs in
utilizing synchronous communication and also by providing a fully
distributed rollback protocol.

\subsection{User-Level Syntax} 

We begin with a core calculus which is rich enough to express the
semantic notions of interest:
\[\begin{array}{lrcl}
(\textit{channel names}) & \ell \\
(\textit{constants}) & c &::=& 
  () \alt 0 \alt 1 \alt \ldots \alt + \alt - \alt \geq \alt \ldots \\
(\textit{expressions}) & e &::=& 
    c \alt x \alt \lambda x.e \alt e_1e_2 \alt 
    \send{\ell}{e} \alt \recvo{x}{\ell}{e} \alt \stable{e} \alt \backtrack~e \\
(\textit{processes}) & p &::=& 
  \parc{p_1}{p_2} ~~\alt~~ \langle e \rangle
\end{array}\]

\noindent A program is a collection of processes executing in
parallel.  Expressions extend the call-by-value $\lambda$-calculus
with communication and backtracking primitives. The communication
primitives are $\send{\ell}{e}$ which commits to sending the value
of~$e$ on the channel $\ell$ and $\recvo{x}{\ell}{e}$ which blocks
until it receives~$x$ on channel $\ell$.  Our Scala implementation
supports input ``choice'' allowing a receiving process to
non-deterministically choose among a collection of active channels;
the addition of choice is necessary for expressiveness but adds little
new insight to the formal semantics and is hence omitted in the interest
of brevity.  The backtracking primitives are $\stable{e}$ which is
used to delimit the scope of possible backtracking events within
$e$. The expression $\backtrack~e$ typically has two effects: the
control state in the process executing the instruction jumps back to
the dynamically closest nested block with the value of~$e$ \emph{and}
all intervening communication events are invalidated. The latter
action might force neighboring processes to also backtrack, possibly
resulting in a cascade of backtracking for a poorly written program.

\subsection{Internal Syntax} 
\label{sec:internalsyntax}
 
In order to formalize evaluation, we define a few auxiliary syntactic
categories that are used to model run-time data structures and
internal states used by the distributed reversible protocol. These
additional categories include process names, time stamps, channel
maps, evaluation contexts, and stacks:

\[\begin{array}{lrcl}
(\textit{process names}) & n \\
(\textit{time stamps}) & t \\
(\textit{values}) & v &::=& c \alt x \alt \lambda x.e \alt \stable{(\lambda x.e)} \\
(\textit{expressions}) & e &::=&  \ldots \alt \stableu{e} \\
(\textit{evaluation contexts}) & E &::=& 
  \Box \alt E~e \alt v~E \alt \send{\ell}{E} \alt \stable{E} \alt \stableu{E} 
  \alt \backtrack~E \\
(\textit{channel maps}) & \Xi &=& \ell \mapsto (n,t,n) \\
(\textit{stacks}) & \Gamma &=& \bullet \alt \Gamma,(E,v,t,\Xi) \\
(\textit{processes}) & p &::=& 
  \parc{p_1}{p_2} ~~\alt~~ \hllproc{n}{t}{\Gamma}{e} \\
(\textit{configurations}) & C &::=& \Xi~\fatsemi~ (p_1 \,\|\, p_2 ~\ldots~ \,\|\, p_k)
\end{array}\]

Expressions are extended with $\stableu{e}$ which indicates an
\emph{active} region. The syntax of processes
$\hllproc{n}{t}{\Gamma}{e}$ is extended to record additional
information: a process id $n$, a virtual time $t$, a context stack
$\Gamma$, and an expression $e$ to evaluate. The processes communicate
using channels $\ell$ whose state is maintained in maps $\Xi$. Each
entry in $\Xi$ maps a channel to the sender and receiver processes
(which are fixed throughout the lifetime of the channel) and the
current virtual time of the channel.  Contexts are pushed on the stack
when a process enters a new stable region and popped when a process
backtracks or exits a stable region. Each context includes a
conventional continuation (modeled by an evaluation context $E$), a
value $v$ with which to backtrack if needed, a time stamp, and a local
channel map describing the state of the communication channels at the
time of the checkpoint.

A semantic configuration $C$ consists of a global channel map $\Xi$
and a number of processes. An invariant maintained by the semantics is
that a process executing in the forward direction will have the times
of its channels in the global map greater than or equal to the times
associated with the channels in the top stack frame.  Similarly, the
time associated with a process will always be at least as great as the
times associated with its channels.  Both invariants follow from the
intuition that any channel appearing on the stack must have been
pushed ``in the past'' and similarly that any communication reflected
in the global map must have also happened ``in the past.'' A process
in the backtracking state will temporarily violate these invariants
until it negotiates a consistent state with its neighbors.

We assume that in the initial system state, all channels have time 0
and every process is of the form
$\hllproc{n}{0}{\bullet}{\stable{(\lambda\_.e)}~()}$; i.e.\ process
$n$ is entering a stable region containing the expression $e$ with an
empty context stack at time 0.

\begin{figure}
\footnotesize{
\begin{align}
& \tag{H1}\label{eq:eval}
  \Xi\fatsemi
  {\hllproc{n}{t}{\Gamma}{E[\diff{(\lambda x.e)~v}]}}
\xrightarrow{\epsilon}
  \Xi\fatsemi
  {\hllproc{n}{t}{\Gamma}{E[\diff{e[v/x]}]}}
\\
\tag{H2} \label{eq:fwdcomm}
\begin{split}
&  \Xi\{\ell\mapsto(n_1,\diff{t_c},n_2)\}\fatsemi 
  \parc{\hllproc{n_1}{\diff{t_1}}{\Gamma_1}{E_1[\diff{\send{\ell}{v}}]}\,\,\,}{
        \hllproc{n_2}{\diff{t_2}}{\Gamma_2}{E_2[\diff{\recvo{x}{\ell}{e}}]}}
  \xrightarrow{\ell@t[v]}  \\
&  \Xi\{\ell\mapsto(n_1,\diff{t},n_2)\}~\fatsemi 
  \parc{\hllproc{n_1}{\diff{t}\,\,}{\Gamma_1}{E_1[\diff{()}]}}{
        \hllproc{n_2}{\diff{t}}{\Gamma_2}{E_2[\diff{e[v/x]}]}} \\
& \text{where}~t > \max(t_1,t_2)  
\end{split}
\\
\tag{H3}\label{eq:push}
\begin{split}
&  \Xi\fatsemi
  {\hllproc{n}{\diff{t}}{\Gamma}
  {E[\diff{(\stable{(\lambda x.e)})~v}]}}
  \xrightarrow{\epsilon} \\
&  \Xi\fatsemi
  {\hllproc{n}{\diff{t'}}
    {\Gamma\diff{,(E[(\stable{(\lambda x.e)})~\Box],v,t,\Xi_n)}}
    {E[\diff{\stableu{e[v/x]}}]}} \\
& \text{where}~t' > t~\text{and}~\Xi_n~\text{is the subset of}~\Xi~
  \text{referring to the channels of}~n
\end{split}
\\
& \tag{H4}\label{eq:push2}
  \Xi\fatsemi
  {\hllproc{n}{t}{\Gamma\diff{,(E',v',t',\Xi')}}{E[\diff{\stableu{v}}]}}
\xrightarrow{\epsilon}
  \Xi\fatsemi
  {\hllproc{n}{t}{\Gamma}{E[\diff{v}]}}
\\
& \tag{H5}\label{eq:backcomm1}
  \Xi\fatsemi\hllproc{n}{t}{\Gamma,(E',v',t',e')}{\diff{e}}
\xrightarrow{\epsilon}
  \Xi\fatsemi\hllproc{n}{t}{\Gamma,(E',v',t',e')}{\diff{\backtrack~v'}}
\\
\tag{H6}\label{eq:backcomm}
\begin{split}
&  \Xi\{\ell\mapsto(c_1,\diff{t_c},c_2)\}\fatsemi 
  \parc{\hllproc{n_1}{t_1}{\Gamma_1}{E_1[\backtrack~v_1]}}
          {\hllproc{n_2}{t_2}{\Gamma_2}{E_2[\backtrack~v_2]}}
  \xrightarrow{\overline{\ell}@t_c'} \\
&  \Xi\{\ell\mapsto(c_1,\diff{t_c'},c_2)\}\fatsemi 
  \parc{\hllproc{n_1}{t_1}{\Gamma_1}{E_1[\backtrack~v_1]}}
          {\hllproc{n_2}{t_2}{\Gamma_2}{E_2[\backtrack~v_2]}} \\
& \text{where}~0 \leq t_c' < t_c~\text{and}~\{ c_1,c_2 \} = \{ n_1, n_2\}
\end{split}
\\
\tag{H7}\label{eq:rebacktrack}
\begin{split}
&     \Xi\{\ell\mapsto(-,t,-)\}\fatsemi
     \hllproc{n_1}{t_1}{\Gamma\diff{,(E_1,v',t_1',\Xi_1\{\ell\mapsto(-,t',-)\})}}{E[\backtrack~v]}
  \xrightarrow{\epsilon} \\
&     \Xi\{\ell\mapsto(-,t,-)\}\fatsemi
     \hllproc{n_1}{t_1}{\Gamma}
     {E[\backtrack~v]} \\
& \text{where}~t < t' 
\end{split}
\\
\tag{H8}\label{eq:reforward}
\begin{split}
&     \Xi\fatsemi
     \hllproc{n_1}{\diff{t_1}}{\Gamma\diff{,(E_1,v_1,t_1',\Xi_1)}}{\diff{E[\backtrack~v]}}
\xrightarrow{\epsilon}
     \Xi\fatsemi
     \hllproc{n_1}{\diff{t_1'}}{\Gamma}{\diff{E_1[v]}} \\
& \text{where the timestamp on every channel in}~\Xi_1~\text{is equal to
  the timestamp of the same channel in}~\Xi
\end{split}
\end{align}
}
\caption{\label{fig:hl}High-level rules}
\end{figure}

\subsection{Forward Semantics} 

The rules are collected in Fig.~\ref{fig:hl}. We discuss each rule
with the aid of simple examples below. 

A computation step that does not involve communication, stable
regions, or backtracking is considered a local computation step. None
of the internal structures need to be consulted or updated during such
local computation steps and hence in our Scala implementation, local
computation steps proceed at ``full native speed.'' In order to
establish notation, here is a simple computation step:\footnote {The
  color version of the paper highlights the components of the
  configuration that are modified by each rule.}

{\footnotesize{
\[
  {\{\ell\mapsto(n_1,2,n_2)\}}\fatsemi \hllproc{n_1}{5}{\Gamma}{\diff{1+2}}
  \rightarrow 
  {\{\ell\mapsto(n_1,2,n_2)\}}\fatsemi \hllproc{n_1}{5}{\Gamma}{\diff{3}}
\]}}

\noindent In this example, a process $n_1$ with local virtual time 5
and making forward progress encounters the computation $1+2$.  We
assume the existence of a channel $\ell$ which is associated with time
2 and connects~$n_1$ to $n_2$. Intuitively this means that the last
communication by that process on that channel happened three virtual
time units in the past. The reduction rule leaves all structures
intact and simply performs the local calculation. In the general case,
we have rule~\ref{eq:eval} for application of $\lambda$-expressions
and similar rules for applying primitive operations. In
rule~\ref{eq:eval}, the runnable expression in the process is
decomposed into an evaluation context $E$ and a current
``instruction'' $(\lambda x.e)~v$. This instruction is performed in
one step that replaces the parameter $x$ with the value $v$ in the
body of the procedure $e$. The notation for this substitution is
$e[v/x]$. The entire transition is tagged with $\epsilon$ indicating
that it produces no visible events.

Communication between processes is synchronous and involves a
handshake.  We require that in addition to the usual exchange of
information between sender and receiver, that the handshake
additionally exchanges several virtual times to force the virtual
times of the sending process, the receiving process, and the used
channel to be all equal to a new virtual time larger than any of
the prior times for these structures. Here is a small example
illustrating this communication handshake:

{\footnotesize{\[\begin{array}{ll}
& {\{\ell\mapsto(n_1,\diff{3},n_2)\}}\fatsemi 
  \hllproc{n_1}{\diff{5}}{\Gamma}{\diff{\send{\ell}{10}}} ~\,\|~
  \hllproc{n_2}{\diff{4}}{\Gamma}{\diff{\recvo{x}{\ell}{x+1}}} \\
\rightarrow & 
  {\{\ell\mapsto(n_1,\diff{6},n_2)\}}\fatsemi 
  \hllproc{n_1}{\diff{6}}{\Gamma}{\diff{()}} \qquad\quad~\, \|~
  \hllproc{n_2}{\diff{6}}{\Gamma}{\diff{10+1}}
\end{array}\]}}

\noindent Initially, we have two processes willing to communicate on channel
$\ell$. Process $n_1$ is sending the value~10 and process $n_2$ is
willing to receive an $x$ on channel $\ell$ and proceed with
$x+1$. After the reduction, the value 10 is exchanged and each process
proceeds to the next step. In addition, the virtual times of the two
processes as well as the virtual time of the channel~$\ell$ have all
been synchronized to time 6 which is greater than any of the previous
times. This is captured in rule~\ref{eq:fwdcomm}. In that rule, the fact that
$t > t_c$ follows from the global model invariant that
$t_1 \ge t_c \wedge t_2 \ge t_c$.  The notation
$\Xi\{\ell\mapsto(n_1,t_c,n_2)\}$ says that, in channel map $\Xi$,
channel~$\ell$ connects sender $n_1$ and receiver~$n_2$ and has time
$t_c$. The transition produces the visible event~$\ell@t[v]$ that
value~$v$ was transferred on channel~$\ell$ at time~$t$.

Finally, there are rules~\ref{eq:push}, \ref{eq:push2},
and~\ref{eq:backcomm1} corresponding to entering and exiting stable
regions.  These rules are interconnected and we illustrate them with a
small example whose first transition is:

{\footnotesize{\[\begin{array}{ll}
& {\{\ell\mapsto(n_1,2,n_2)\}}\fatsemi 
  \hllproc{n_1}{\diff{5}}{\Gamma \qquad\qquad\qquad\qquad\qquad\qquad\qquad\qquad~}
  {7 + \diff{(\stable{f})~v}} \\
\rightarrow & 
  {\{\ell\mapsto(n_1,2,n_2)\}}\fatsemi 
  \hllproc{n_1}{\diff{6}}{\Gamma\diff{(7 + (\stable{f})~\Box,v,5,\{\ell\mapsto(n_1,2,n_2)\})}}
  {7 + \diff{\stableu{(f~v)}}}
\end{array}\]}}

\noindent Process $n_1$ encounters the expression $7 + (\stable{f})~v$ where $f$
is some function and $v$ is some value. The \textsf{stable} construct
indicates that execution might have to revert back to the current
state if any backtracking actions are encountered during the execution
of $f~v$. The first step is to increase the virtual time to establish
a new unique event. Then, to be prepared for the eventuality of
backtracking, process $n_1$ pushes
$(7 + (\stable{f})~\Box, v, 5, \{\ell\mapsto(n_1,2,n_2)\})$ on its
context stack. The pushed information consists of the continuation
$7+\stable{f}~\Box$ which indicates the local control point to jump
back to, the value $v$, the virtual time 5 which indicates the time to which to
return, and the current channel map which captures the state of the
communication channels to be restored. Execution continues with
$7 + \stableu{(f~v)}$ where the underline indicates that the region is
currently active. If the execution of $f~v$ finishes normally, for
example, by performing communication on channel $\ell$ and then
returning the value 100, then the evaluation progresses as follows:

{\footnotesize{\[\begin{array}{ll}
& {\{\ell\mapsto(n_1,8,n_2)\}}\fatsemi 
  \hllproc{n_1}{8}{\Gamma\diff{,(7 + (\stable{f})~\Box,v,5,\{\ell\mapsto(n_1,2,n_2)\})}\,\,}
  {7 + \diff{\stableu{100}}} \\
  \rightarrow & 
 {\{\ell\mapsto(n_1,8,n_2)\}}\fatsemi 
  \hllproc{n_1}{8}{\Gamma \qquad\qquad\qquad\qquad\qquad\qquad\qquad\qquad~~~\,}
  {7 + \diff{100}}
\end{array}\]}}

\noindent The context stack is popped and execution continues in the
forward direction. The case in which $f~v$ backtracks is considered in
the next section.

\subsection{Backtracking Semantics} 

Backtracking may occur either from within the current process or
indirectly because another neighboring process has retracted a
communication event. In the first case, the saved context will be
resumed with a value of the programmer's choice; in the latter case,
the context will be resumed, asynchronously, with the value saved on
the context stack. A well-typed program should have the function
argument to \textsf{stable} ($\lambda x.e$ in the rules above) be
prepared to handle either situation.

We illustrate the important steps taken in a typical backtracking
sequence using an example. Consider the following configuration in
which process $n_1$ has communicated on channel $\ell$ at time 2,
taken a step to time 3, entered a stable region, communicated again on
$\ell$ at time 5, taken two steps to time 7, entered another stable
region, communicated again on $\ell$ at time 8, taken a step to time
9, and then encountered a backtracking instruction. The internal state
of its communicating partner is irrelevant for the example except that
its virtual time is assumed to be larger than 8:

{\footnotesize{\begin{tabbing}
     $\qquad$\=$\{\ell\mapsto(n_1,\diff{8},n_2)\}$\=$\fatsemi\langle$\=$n_1@9: \Gamma$\=$
    ,(E_1[(\stable{f_1})~\Box],v_1,3,\{\ell\mapsto(n_1,2,n_2)\})$  \\
  \> \>\>\>$,(E_2[(\stable{f_2})~\Box],v_2,7,\{\ell\mapsto(n_1,5,n_2)\}),$ \\
  \> \>\> ${~E_3[\backtrack~100]}\rangle$ \\
  \> \> $\,\|\, \langle n_2@13: \Gamma_2,e_2 \rangle$
\end{tabbing}}}

\noindent The first step is for two processes to negotiate a time to
which channel $\ell$ should return. The semantic specification is
flexible allowing \emph{any} time in the past (including 0 in the
extreme case). For the running example, we pick time 2 for the channel
$\ell$. The configuration steps to:

{\footnotesize{\begin{tabbing}
     $\qquad$\=$\{\ell\mapsto(n_1,\diff{2},n_2)\}$\=$\fatsemi\langle$\=$n_1@9: \Gamma$\=$
    ,(E_1[(\stable{f_1})~\Box],v_1,3,\{\ell\mapsto(n_1,2,n_2)\})$  \\
  \> \>\>\>$,(E_2[(\stable{f_2})~\Box],v_2,7,\{\ell\mapsto(n_1,5,n_2)\}),$ \\
  \> \>\> ${~E_3[\backtrack~100]}\rangle$ \\
  \> \> $\,\|\, \langle n_2@13: \Gamma_2,e_2 \rangle$
\end{tabbing}}}

\noindent At this point, neither process may engage in any forward
steps. Focusing on $n_1$ for the remaining of the discussion, the top
stack frame needs to be popped as its embedded time for channel $\ell$
is \emph{later} than the global time:

{\footnotesize{\begin{tabbing}
$\qquad$\=$\{\ell\mapsto(n_1,2,n_2)\}\fatsemi\langle$\=$n_1@9: \Gamma$\=$
    ,(E_1[(\stable{f_1})~\Box],v_1,3,\{\ell\mapsto(n_1,2,n_2)\}),$
   ${E_3[\backtrack~100]}\rangle$ 
\end{tabbing}}}

\noindent In general, the popping of stack frames continues until the
time associated with all the channels in the top stack frame agrees
with the global times associated with the channels. This is guaranteed
to be satisfied when the process backtracks to the initial state but
might, as in the current example, be satisfied earlier. In this case,
forward execution resumes with the stable region saved in the top
stack frame:

{\footnotesize{\begin{tabbing}
$\qquad$\=
  $\{\ell\mapsto(n_1,2,n_2)\}\fatsemi\langle$\=$n_1@3: \Gamma,$
  ${E_1[(\stable{f_1})~100]}\rangle$ 
\end{tabbing}}}

\noindent As illustrated in the example above, our semantic rules impose as few
constraints as possible on the extent of, and the number of steps
taken during backtracking, to serve as a general specification; our
Scala implementation constrains the application of these rules to
obtain an efficient implementation.  

Formally, we have four rules. The rule~\ref{eq:backcomm1} allows a
process to asynchronously enter a backtracking
state. Rule~\ref{eq:backcomm} allows a pair of communicating processes
in the backtracking state (either because they encountered the
backtracking command itself or asynchronously decided to backtrack
using the rule above) to select any earlier time for their common
channel. The label $\overline{\ell}@t_c'$ means that all communication
events on channel~$\ell$ at times later than $t_c'$ are retracted.
Notice that only the channel time is reduced -- this preserves our
invariant that the virtual time of every process is greater than or
equal to that of its channels.
Rule~\ref{eq:rebacktrack} allows a process in the backtracking state
to pop stack frames that were pushed after the required reset time of
the channel. Finally, in rule~\ref{eq:reforward}, a process that is
backtracking can return to forward action if all its channels are in a
``consistent'' state -- that is, when all channels in its stored
channel map in the top stack frame have timestamps matching what is
found in the global channel map.  It is only at this point that the
virtual time of the process is updated.

\section{Channel Protocol}
\label{sec:chanprotocol}

The high-level semantics assumes that synchronous communication is
realized atomically (e.g., rule H2). In an actual implementation,
synchronization between sender and receiver requires a multi-phase
handshaking protocol. It is not evident that such low-level protocols
are robust if interleaved with backtracking actions. Because of the
subtlety of this point, we formalize a low-level communication
protocol with backtracking and prove it correct. In the next section,
we will introduce a low-level semantics and use invariants of the
protocol to prove that the high-level semantics can be faithfully
implemented without assuming atomic synchronous communication.

\subsection{Low-Level Communication}
 
Instead of assuming that synchronous communication is atomic, we
consider instead the realistic situation in which communication
happens in two phases: (i) a sender requests a communication event,
and (ii) after some unspecified time the receiver acknowledges the
request.  

In a distributed environment, where channel state changes made by the
sender or receiver take time to propagate, the low-level communication
messages introduce potential race conditions. We account for race
conditions by verifying a model where communication is buffered. Thus,
global channel state will be divided into two parts -- one maintained
by the sender and the other maintained by the receiver.  A process may
only write the state associated with its channel end, and may only
read a delayed version of the state maintained by its communicating
partner. This requires that channels carry two timestamps -- one
maintained by the sender and one maintained by the receiver.  We 
think of the timestamp maintained by the receiver as the ``true''
channel time. In addition to independent timestamps, each end of the
channel will also have a token bit and a ``direction'' flag.  The
token bits jointly determine which end of the channel may make the
next ``move,'' and the flag (loosely) determines the direction of
communication, forward or backwards.

A crucial aspect exposed by the low-level communication protocol is
the ability of a blocked sender or receiver engaged in a
synchronization to signal its partner that it wishes to switch from
forward to backwards communication. To accommodate such situations,
the receiver state will also include an auxiliary Boolean variable
\lstinline|sync| that is set when the receiver agrees to complete a
communication event and reset when it refuses a communication
event. This variable is not visible to the sender; however, the sender
will be able infer its value from the visible state even in the
presence of potential races.

\subsection{Protocol Types}

We now introduce the formal model for the channel protocol using the
Symbolic Analysis Laboratory (SAL)
tools~\cite{Moura03,Moura04,ds04}. In the SAL language the key
protocol types are defined as:

\lstinputlisting[firstline=12,  lastline=13]{\salmodel}

\noindent As motivated above, each of the sender and receiver states
are defined by four state variables:

\lstinputlisting[firstline=17,  lastline=20]{\salmodel}
\lstinputlisting[firstline=68,  lastline=70]{\salmodel}

The state of a channel consists of the union of the sender and
receiver states.  In general, the right to act alternates between the
sender and the receiver.  The sender is permitted to initiate a
communication event (forwards or backwards) when the two token bits
are equal.  The receiver is permitted to complete a communication
event when the two token bits are unequal.  Thus the sender (receiver)
``holds'' the token when these bits are equal (unequal). This
alternating behavior is a characteristic of handshake protocols.

\subsection{Model}

In SAL, transition rules are simply predicates defining pre- and
post-conditions; the next state of \lstinline+s_b+ is \lstinline+s_b'+. Forward
communication is initiated when the sender executes the guarded
transition:

\begin{trans} \label{lst:fwdreq}
  \lstinputlisting[firstline=40,lastline=41]{\salmodel}
\end{trans}

\noindent Thus, the sender may initiate forward communication whenever
it holds the ``token'' (\lstinline+s_b = r_b+) and the receiver is
accepting forward transactions (\lstinline+r_d = F+).  By executing the
transition, the sender selects a new time (\lstinline+s_t'+), relinquishes
the token, indicates that it is executing a forward transaction
(\lstinline+s_d' = F+), and selects (arbitrary) data to transfer.

The receiver completes the handshake by executing the following
transition in which it updates its clock (to a value at least that
offered by the sender), and flips its token bit.  This transition is
only permitted when both the sender and the receiver wish to engage in
forward communication:

\begin{trans} \label{lst:fwdack}
 \lstinputlisting[firstline=88, lastline=89]{\salmodel}
\end{trans}

A receiver may also refuse a forward transaction by indicating that it
desires to engage only in backwards communication:

\begin{trans} \label{lst:fwdnack}
\lstinputlisting[firstline=99,  lastline=99]{\salmodel} 
\end{trans}

Our protocol also supports backwards communication events.  The sender
may initiate a backwards event whenever it holds the token:

\begin{trans} \label{lst:backreq}
\lstinputlisting[firstline=49,  lastline=49]{\salmodel}
\end{trans}

In a manner analogous to forward communication, the receiver may
complete the event by executing the following transition.  One
subtlety of this transition is that the receiver may also signal
whether it is ready to resume forward communication (\lstinline+r_d' = F+)
or wishes to engage in subsequent backward events (\lstinline+r_d'= B+).
The latter occurs when the sender has offered a new time that is not
sufficiently in the past to satisfy the needs of the receiver:

\begin{trans} \label{lst:backack}
\lstinputlisting[firstline=116,  lastline=117]{\salmodel}
\end{trans}

While the protocol presented supports both forward and backwards
communication, the sender may be blocked waiting for a response from a
receiver when it wishes to backtrack.  The following transition allows
the sender to {\it request} that a forward transaction be retracted:

\begin{trans} \label{lst:sendint}
 \lstinputlisting[firstline=57,  lastline=57]{\salmodel}
\end{trans}

The receiver may either accept the original offer to communicate
(\textit{Trans}~\ref{lst:fwdack}) or allow the retraction:

\begin{trans} \label{lst:sendintack}
\lstinputlisting[firstline=107,  lastline=107]{\salmodel}
\end{trans}

Similarly a blocked receiver may signal the sender that it wishes to
backtrack:

\begin{trans} \label{lst:rcvint}
\lstinputlisting[firstline=125,  lastline=125]{\salmodel}
\end{trans}

\subsection{Key Invariant} 

The main subtleties that we need to verify occur when a sender may
attempt to retract a forward request while the receiver simultaneously
acknowledges that request, or when a receiver may decide, after it has
acknowledged a request, that it wishes to backtrack.  In either case
the later decision ``overwrites'' state that may or may not have been
seen by the partner.  For our semantic model, it is crucial that the
sender be able to determine whether the synchronization event occurred
or was successfully retracted.  The key invariant of our SAL model
uses a shadow variable to prove this property.

\section{Low-Level Processes and Refinement Mapping} 
\label{sec:ll}

Our low-level model is derived from the high-level model by
implementing those rules involving synchronization using finer-grained
rules based upon the the protocol model.  Necessarily, there are more
transition rules associated with the low-level model.  For example,
the single high-level transition implementing forward communication
requires three transitions (two internal and one external or visible)
in the low-level model.  High level transitions not involving
communication (\ref{eq:eval}, \ref{eq:push}, \ref{eq:push2}, and
\ref{eq:rebacktrack}) are adopted in the low-level model with minimal
changes to account for the differences in channel state.
The
high level transitions \ref{eq:backcomm1} and \ref{eq:reforward} are
adopted with a few additional side conditions.

Along with our presentation of the low-level model (henceforth \LP),
we sketch a refinement mapping from \LP\ to the high-level model
(henceforth \HP) -- essentially a function that maps every state of
\LP\ to a state of \HP\ and where every transition of \LP\ maps to a
transition (or sequence of transitions) of \HP.
\cite{Abadi:1991:ERM:114015.114018}

\begin{figure}
\footnotesize{
\begin{align}
\label{eq:fwdinit} \tag{L1}
\begin{split}
& \Xi\{\ell\mapsto(s:(n,t_s,b,-,-),r:(n_r,t_r,b,F)\}\fatsemi
  {\hllproc{n}{t}{\Gamma}{E[\send{\ell}{v}]}}
\xrightarrow{\epsilon} \\
& \Xi\{\ell\mapsto(s:(n,\diff{t_s'},\diff{\overline{b}},F,\diff{v}),
  r:(n_r,t_r,b,F)\}\fatsemi
 {\hllproc{n}{t}{\Gamma}{E[\diff{\sendu{\ell}{v}}]}} \\
& \text{where}~t_s' > t_r
\end{split}
\\
\tag{L2}\label{eq:fwdack}
\begin{split}
& \Xi\{\ell\mapsto(s:(n_s,t_s,b,d_s,v),r:(n,t_r,\overline{b},F)\}\fatsemi 
 {\hllproc{n}{t}{\Gamma}{E[\recvo{x}{\ell}{e}]}} 
\xrightarrow{\ell@t[v]} \\
& \Xi\{\ell\mapsto(s:(n_s,t_s,b,d_s,v),r:(n,\diff{t},\diff{b},F)\}\fatsemi
      {\hllproc{n}{t}{\Gamma}{E[\diff{e[v/x]}]}} \\
& \text{where}~t > \max(t_s, t)~\text{and}~d_s \in \{F, I\}
\end{split}
\\
\tag{L3}\label{eq:fwdcomplete}
\begin{split}
& \Xi\{\ell\mapsto(s:(n,t_s,b,d_s,v),r:(n_r,t_r,b,d_r)\}\fatsemi
 {\hllproc{n}{t}{\Gamma}{E[\sendu{\ell}{v}]}}
\xrightarrow{\epsilon} \\
& \Xi\{\ell\mapsto(s:(n,t_s,b,d_s,v),r:(n_r,t_r,b,d_r)\}\fatsemi
 {\hllproc{n}{\diff{t_r}}{\Gamma}{E[\diff{()}]}} \\
& \text{where}~d_s \ne B~\text{and}~t_s \le t_r
\end{split}
\\
\tag{L4}\label{eq:fwdnack}
\begin{split}
& \Xi\{\ell\mapsto(s:(n_s,t_s,b,F,v),r:(n,t_r,\overline{b},-)\}\fatsemi
                {\hllproc{n}{t}{\Gamma}{E[\backtrack~v]}}
                \xrightarrow{\epsilon} \\
&                 \Xi\{\ell\mapsto(s:(n_s,t_s,b,F,v),r:(n,t_r,\diff{b},B)\}\fatsemi
                 {\hllproc{n}{t}{\Gamma}{E[\backtrack~v]}} \\
& \text{where}~t_r > 0
\end{split}
\\
\tag{L5}\label{eq:backinit}
\begin{split}
&  \Xi\{\ell\mapsto(s:(n,t_s,b,-,v),r:(n_r,t_r,b,d)\}\fatsemi
                  {\hllproc{n}{t}{\Gamma}{E[\backtrack~v]}}
                  \xrightarrow{\epsilon} \\
&                  \Xi\{\ell\mapsto(s:(n,\diff{t'_s,
                    \overline{b},B},v), r:(n_r,t_r,b,d)\}\fatsemi
                     {\hllproc{n}{t}{\Gamma}{E[\backtrack~v]}} \\
& \text{where}~t'_s < t_r
\end{split}
\\
\tag{L6}\label{eq:backack}
\begin{split}
&  \Xi\{\ell\mapsto(s:(n_s,t_s,\overline{b},B,v),r:(n,t,b,-)\}\fatsemi
                  {\hllproc{n}{t}{\Gamma}{E[\backtrack~v]}}
                  \xrightarrow{\overline{\ell}@t_s} \\
&                  \Xi\{\ell\mapsto(s:(n_s,t_s,\overline{b},B,v),
                  r:(n,t_s,\diff{\overline{b}},\diff{F})\}\fatsemi
                  {\hllproc{n}{t}{\Gamma}{E[\backtrack~v]}} 
\end{split}
\\
\tag{L7} \label{eq:rcvsignal}
\begin{split}
&  \Xi\{\ell\mapsto((n_s,t_s,b,d_s,v),(n,t_r,b,F)\}\fatsemi
                  {\hllproc{n}{t}{\Gamma}{E[\backtrack~v]}}
                  \xrightarrow{\epsilon} \\
&                  \Xi\{\ell\mapsto((n_s,t_s,b,d_s,v),(n,t_r,b,\diff{B})\}\fatsemi
                     {\hllproc{n}{t}{\Gamma}{E[\backtrack~v]}} \\
& \text{where}~t_r > 0
\end{split}
\\
\tag{L8}\label{eq:retractreq}
\begin{split}
&  \Xi\{\ell\mapsto(s:(n,t_s,\overline{b},F,v),r:(n_r,t_r,b,d_r)\}\fatsemi
                  {\hllproc{n}{t}{\Gamma}{E[\sendu{\ell}{v}]}}
                  \xrightarrow{\epsilon} \\
&                  \Xi\{\ell\mapsto(s:(n,t_s,{\overline{b}},\diff{I},v),
                  r:(n_r,t_r,b,d_r)\}\fatsemi
                  {\hllproc{n}{t}{\Gamma}{E[\sendu{\ell}{v}]}}
\end{split}
\\
\tag{L9}\label{eq:retractack}
\begin{split}
&  \Xi\{\ell\mapsto((n_s,t_s,\overline{b},I,v),(n,t_r,b,d)\}\fatsemi
                  {\hllproc{n}{t}{\Gamma}{E[e]}}
                  \xrightarrow{\epsilon} \\
&                  \Xi\{\ell\mapsto((n_s,t_s,\overline{b},I,v),(n,t_r,\diff{\overline{b}},\diff{d'})\}\fatsemi
                     {\hllproc{n}{t}{\Gamma}{E[e]}} \\
& \text{where} ~d' \in \{d, B\}
\end{split}
\\
\tag{L10}\label{eq:retractcomp}
\begin{split}
&  \Xi\{\ell\mapsto((n,t_s,b,I,v),(n_r,t_r,b,d_r)\}\fatsemi
  \hllproc{n}{t}{\Gamma}{E[\sendu{\ell}{v}]} 
   \xrightarrow{\epsilon} \\
&  \Xi\{\ell\mapsto((n,t_s,b,I,v),(n_r,t_r,b,d_r)\}\fatsemi
  \hllproc{n}{t}{\Gamma}{E[\diff{\send{\ell}{v}}]} \\
& \text{where}~t_r > t_s 
\end{split}
\\
\tag{H8 Condition}\label{eq:lowreforward}
& (\Xi\{\ell\mapsto(s:(n_1,-,b,-),r:(-,t,b,-))\} \ \vee
   \ \Xi\{\ell\mapsto(s:(-,-,-,-),r:(n_1,t,-,F))\}) 
\end{align}
}
\caption{\label{fig:ll}Low-level rules}
\end{figure}

\subsection{Low-Level Synchronization}

The low-level model is defined by its own set of transition rules
(Figure~\ref{fig:ll}) along with additional state information relating
to the channel implementation.  We begin by discussing in those
transitions relating to forward communication.  We use these
transitions to illustrate how low-level events ``map'' to high-level
events.  The high-level semantics includes rule \ref{eq:fwdcomm}
(forward communication) that requires simultaneous changes in two
processes and rule \ref{eq:backcomm} that depends upon the state of
two processes; none of the low-level rules affects more than one
process and the only preconditions on any of our low-level rules are
the state of a single process and the state of its channels.
Furthermore, these rules modify only the process state and the portion
of a channel state (send or receive) owned by the process.

We define the state of a channel as a tuple
$(s:(n,t,b,d,v),r:(n,t,b,d))$ where $s$ is the state of the sender and
$r$ is the state of the receiver; $s.n$ is the sender id and $r.n$ is
the receiver id. As mentioned, the sender and receiver both maintain
(non-negative) timestamps ($s.t$, $r.t$) and Boolean tokens ($s.b$,
$r.b$).  Each also maintains a direction flag ($s.d$, $r.d$)
indicating ``forward'' or ``backward'' synchronization.  The sender
state includes a value $s.v$ to be transferred when communication
occurs. These state elements correspond to those of the channel
protocol.  The refinement mapping from \LP\ to \HP\ drops this additional
channel state information; although it does impact the mapping of
the process expressions.



Forward communication (\ref{eq:fwdcomm} in \HP) is executed in three
steps by the underlying channel protocol.

\begin{itemize}
\item[L1] In the first step, which corresponds to
  \textit{Trans}~\ref{lst:fwdreq} of the SAL model, the sender
  initiates the communication by marking its state as ``in progress''
  with the new expression $\sendu{\ell}{v}$. This expression has no
  direct equivalent in \HP and may only occur as a result of this
  rule.

Recall that the sender has the ``token'' when the two channel token
bits are equal ($s.b = r.b$), and initiates communication by
inverting its token bit ($s.b$).





\item[L2] In the second step, (\textit{Trans}~\ref{lst:fwdack}) the
  receiver ``sees'' that the sender has initiated communication, reads
  the data, updates its local virtual time, updates the channel's
  time, and flips its token bit to enable the sender to take the next
  and final step in the communication. (Note that the sender stays
  blocked until the receiver takes this step.) After taking this step
  the receiver can proceed with its execution:

From \ref{eq:fwdinit} we can show that $t_s > t_r$.  A required
invariant for our model is that when the conditions of this rule are
satisfied, the sender $n_s$ is executing $\sendu{\ell}$.

\item[L3] In the final step, the sender notes that its active
  communication event has been acknowledged by the receiver. It
  updates its local time and unblocks.

\end{itemize}

The introduction of new control states such as $\sendu{\ell}{v}$
necessarily complicates the creation of a refinement mapping, which
maps this to either $\send{\ell}{v}$ or $()$ depending upon the
state of the channel.  This follows naturally from the protocol
in which a sender initiates a synchronization event, but the receiver
completes it.
Consider the following
cases for mapping of $\sendu{\ell}{v}$.  The first case corresponds to
the state after transition \ref{eq:fwdinit} and the second to the
state after transition \ref{eq:fwdack}.  (Recall that $\ell$ is a
channel, and $\ell.x.y$ correspond to fields of the channel state).

\vspace{-1em}{\footnotesize{
\begin{align*}
f(E[{\sendu{\ell}{v}}])
&=  E[{\send{\ell}{v}}]
\ \texttt{if}\ (\ell.s.b \ne \ell.r.b) \vee \ell.s.t > \ell.r.t\\ 
f(E[{\sendu{\ell}{v}}])
&=  E[()]
\ \texttt{if}\ (\ell.s.b  = \ell.r.b) \wedge \ell.s.t \le \ell.r.t \\
\end{align*}}}

\vspace{-3em} \noindent Thus in the mapping, \ref{eq:fwdinit} is a
``silent'' (stuttering) transition and \ref{eq:fwdack}, which is only
executed in parallel with a process executing $\sendu{\ell}{v}$,
corresponds to high-level transition \ref{eq:fwdcomm}.  Notice that
the mapping of this control state depends upon the state of channel
$\ell$, although the mapping drops this additional state information
from the channels.  Finally, \ref{eq:fwdcomplete} maps to \HP\ as a
silent transition.

\subsection{Backtracking Communication}

Backwards communication is considerably more complex for several
reasons.  First, consider that a process wishing to backtrack may
communicate with any of its peers -- at the low level this may involve
simultaneous communication along its various channels.  A further
complication has to do with the fact that low-level communication
involves a handshake between the sender and receiver where the sender
``commits'' and then the receiver may acknowledge.  If a sender has
committed on a channel and detects, through its other channels, that a
peer wishes to backtrack it must somehow retract the outstanding
communication event.

As with forward communication, many of the transitions relating
to backwards communication implement specific transitions in the channel
model.

\begin{itemize}

\item[\ref{eq:backinit}] A sender may initiate a backwards event
  (\ref{eq:backcomm} in \HP) if it is in the backtracking state and
  ``holds'' the token (\textit{Trans}~\ref{lst:backreq}).

Notice that the sender's time did not change; the
time will change when the sender transitions from backwards to
forwards operation (see \ref{eq:reforward})

\item[\ref{eq:backack}] A backtracking channel receiver may
  acknowledge the backwards transaction
  (\textit{Trans}~\ref{lst:backack}).

\item[\ref{eq:rcvsignal}] A receiver that is backtracking may need to
  signal a sender that it wishes to backtrack
  (\textit{Trans}~\ref{lst:rcvint}).

Notice this simply raises a flag requesting a
backtracking event.  It is up to the sender to pay attention by
requesting a backwards communication event (after
entering the backtracking state if necessary).

\item[\ref{eq:fwdnack}] Finally, a receiver may reject a request for
  forward communication (this should only occur if the receiver is
  backtracking) (\textit{Trans}~\ref{lst:fwdnack}).  Notice that in
  this case $t_s > t_r$.

\end{itemize}

\subsubsection{Retracting Forward Requests}

As discussed in Sec.~\ref{sec:chanprotocol}, we have special
actions that allow a blocked sender or receiver to request backward
communication.

\begin{itemize}

\item[\ref{eq:retractreq}] If the sender is blocked, it may ask the
  receiver to allow it to retract the communication request -- this
  should only be executed where the sender has been requested to
  backtrack through some other channel
  (\textit{Trans}~\ref{lst:sendint}).

\item[\ref{eq:retractack}] The receiver {\em may} allow the sender to
  retract the communication (\textit{Trans}~\ref{lst:sendintack}) or
  it may acknowledge the communication using the rule~\ref{eq:fwdack}
  above.  We have a proof obligation to show that the $n_s$ is
  executing $\sendu{\ell}{v}$ whenever $\ell.s.d = I$.  Furthermore,
  we have an obligation to show that $t_s > t_r$.

\item[\ref{eq:retractcomp}]
Once a sender has been permitted to retract its request, it may return
to the sending state from which it is free to respond to requests to
backtrack.  However, it is important to note that the receiver {\it
  may} have completed the request (\ref{eq:fwdcomplete}).  These two
cases are covered by the invariant in our SAL
model. 
\end{itemize}

\subsubsection{Initiating and Exiting Backtracking}

There are two ways a process can begin backtracking -- explicitly
through a $\backtrack$ command in the program text or
spontaneously. The low-level semantic rule for explicitly entering the
backtracking state is the same as the high level rule
(\ref{eq:backcomm1}).  However, because communication involves a series
of protocol steps, we restrict spontaneous backtracking to the case
where a process has no outstanding send request (i.e.\ it is not in
the $\underline{\textsf{send}}$ state).\footnote{Our Scala
  implementation further restricts spontaneous backtracking
  transitions to situations where a neighbor is backtracking; however,
  that does not alter the underlying model.}  Note that the
rule~\ref{eq:retractreq} allows a blocked sender to request retraction
of an outstanding request and hence transition to a state where this side
condition is satisfied.









Finally, we need a rule that allows a process to exit the backtracking
state -- \ref{eq:reforward} with constraints.
The conditions on timestamps are the same, but send channels are
required to be in a quiescent state.  Since the receiver can only
change the timestamp when it has the token (non-quiescent state), this
predicate can safely be evaluated sequentially. For process $n_1$ and
channel $l$ this side condition is (\ref{eq:lowreforward}) in the
figure.  Although checking this condition requires testing the state
of all the process' channels, the channel protocol guarantees that if
a channel satisfies the necessary condition then that property is
stable.  Note that for send channels we require that the sender holds
the token, and receive channels we require that the process has not
requested further backtracking (this may occur in the model, with its
non-determinism, but does not occur in our implementation).

\section{Conclusion}

We have introduced a CSP based language supporting reversible
distributed computing along with two semantic models -- a high-level
model in which synchronous events are modeled by transitions that
affect two processes simultaneously, and a low-level model in which
transitions affect a single process.  These two models are related by
a verified communication protocol which is the basis for the finer
grained transitions of the low-level model.  We outlined a refinement
mapping that we developed proving that the low-level model implements
the high-level model.  This proof required the invariants of the
protocol that were verified with the SAL model checker.  We have also
proved that the high-level model obeys sensible causal ordering
properties even in the face of backtracking.

While our Scala language implementation is somewhat richer than the
simple models presented here (e.g.,\ it supports communication choice,
and dynamic process and channel creation); at its core it is
implemented exactly as indicated by our low-level model.  Channels are
implemented via message passing where the messages carry the channel
state of our protocol.  Processes are implemented as Java threads.
Processes learn that their peers wish to backtrack by examining the
(local) state of their channels.
Stable sections consist of: saving the channel timestamps on the
context stack, executing the stable code in a try/catch block, and
popping the stack; backtracking is implemented by throwing an
exception.  The implementation required approximately 1200 lines of
code.\footnote{Download:
  \href{http://cs.indiana.edu/~geobrown/places-code.tar.gz}
  {http://cs.indiana.edu/$\sim$geobrown/places-code.tar.gz}.}


This paper provides clear evidence that implementing reversible
communicating processes in a distributed manner is both feasible and,
from the perspective of communication overhead, relatively efficient.
We note that our high-level model is somewhat unsatisfying because it
exposes the programmer to the mechanics of backtracking.  In our
current model, even when a process has decided that it wishes to
backtrack, its peers may continue forward execution for a period
during which they may learn from their peers.  If we were to restrict
our attention to traditional roll-back recovery, where nothing is
``learned'' from unsuccessful forward execution, this could easily be
abstracted.  We continue to work towards a ``compromise'' between
traditional rollback and the unrestricted model we have presented.

\paragraph*{Acknowledgment.} 

This material is based upon work supported by the National Science
Foundation under grant numbers 1116725 and 1217454 and while the first
author was an NSF program officer.  Any opinion, findings, and
conclusions or recommendations expressed in this material are those of
the authors and do not necessarily reflect the views of the National
Science Foundation.


\bibliographystyle{eptcs}
\bibliography{p}
\end{document}